# Quantum entanglement based on surface phonon polaritons in condensed matter systems

**Yang Ming, Zi-jian Wu, Xi-kui Hu, Fei Xu* and Yan-qing Lu***

*National Laboratory of Solid State Microstructures and College of Engineering and Applied Sciences, Nanjing University, Nanjing 210093, China*

*[*]Corresponding authors: yqlu@nju.edu.cn, feixu@nju.edu.cn*

Condensed matter systems are potential candidates to realize the integration of quantum information circuits. Surface phonon polariton (SPhP) is a special propagation mode in condensed matter systems. We present an investigation on the entanglement of SPhP modes. The entangled pairs are generated from entangled photons injected to the system. Quantum performances of entangled SPhPs are investigated by using the interaction Hamiltonian and the perturbation theory. The wave mechanics approach is taken to describe the coupling process as a comparison. Finally, the correlation of system is examined. A whole set of descriptions of SPhP entanglement thus are presented.

Quantum entanglement has become one of the most attractive topics during the past two decades. The quantum entanglement systems are described by entangled states that could not be decomposed into products of single particle states[1]. Due to the development of experimental technologies in recent years, quantum entanglement has been used in varieties of domains, such as quantum cryptography[2], quantum imaging[3], and quantum computation[4]. Usually, the entanglement property of the multiparticle system is demonstrated by entangled photon states[5], for the reason that coupling between photon and environment is relatively small, the decoherence time of the entangled system thus is longer than that of the electrons based systems. However, there are still many rooms to improve in photon systems. For example, bulk optical elements are normally used, which confines the flexibility and scalability of the photon system in practical applications.

To seek for a more practical entanglement system, kinds of improving schemes have been proposed. For example, the usually used spontaneous parametric down-conversion (SPDC) source of entangled photon pairs could be substituted by a domain-engineered nonlinear photonic crystal which integrates the function of some optical elements together[6]. Entanglement systems based on metallic microstructure devices which could support surface plasmon polariton (SPP) mode are also investigated for effective miniaturization[7, 8], as the SPP has a unique capability to confine the electromagnetic field in the perpendicular direction. Recently, owing to meaningful distinct characters, entanglement of phonon modes in condensed matter systems also attracts plenty of attention, and some relative interesting results have been reported[9, 10].

In this letter, we present an investigation of quantum entanglement in condensed matter systems. Unlike the traditional ones[9, 10] in which the entangled phonon states are generated by Raman scattering, the entangled states which we investigate in this work are surface phonon polariton (SPhP) pairs. The SPhP mode is a transverse magnetic (TM) mode vibration resulting from the coupling of an infrared photon (TM mode) with a transverse-optic (TO) phonon[11, 12]. For classical phonon modes in condensed matter systems, thermal fluctuation usually causes serious interference in measurement and destroys the quantum coherence [10]. As an advantage, SPhP does not face this difficulty because it could be regarded as a surface electromagnetic wave in a general view[13]. Moreover, properties of SPhP are analogous to those of surface plasma polariton, so subwavelength-scale quantum circuits also could be realized in condensed matter systems of polar dielectric materials. Overall, SPhP entangled pairs are very suitable candidates for investigations of quantum entanglement in condensed matter systems.

For a lossy, linear, isotropic polar dielectric material, the dispersion relation of

SPhP mode is shown in the following formulation[14]

$$\beta^2 = \frac{\omega^2}{c^2}\left(\frac{\varepsilon_d \varepsilon_c}{\varepsilon_d + \varepsilon_c}\right), \tag{1}$$

where β is the propagation constant of the surface mode. $\varepsilon_d$ and $\varepsilon_c$ are dielectric constants of the two materials located at both sides of the interface, respectively. $\varepsilon_c$ represents the relative permittivity of the lossy polar dielectric material that could be given by the formulation[15]

$$\varepsilon_c(\omega) = \varepsilon_c(\infty)\left[1 + \frac{\omega_{LO}^2 - \omega_{TO}^2}{\omega_{TO}^2 - \omega^2 - i\gamma\omega}\right]. \tag{2}$$

$\omega_{LO}$ and $\omega_{TO}$ are the frequencies of the longitudinal optical phonon and the transverse optical phonon, respectively. And the term $i\gamma\omega$ represents the damping. Usually, γ is much smaller than ω; so for frequencies falling between the TO and LO frequencies, the dielectric constant have a negative real part and a positive imaginary part. Thus the propagation constant also shows a similar form, just like that of the surface plasmon polariton. For a detailed illustration, we take SiC system as an example. Corresponding dielectric permittivity and dispersion relation of SiC are shown in Fig. 1(a) and Fig. 1(b), respectively. Other parameters are given as $\omega_L$=969cm$^{-1}$, $\omega_T$=793cm$^{-1}$, Γ=4.76cm$^{-1}$, $\varepsilon_\infty$=6.7.[14]

As a derivation, when the thickness of the condensed matter becomes thinner enough, the SPhPs supported by the top and bottom surfaces will couple with each other to form a long range surface phonon polariton (LRSPhP) mode. The LRSPhP mode experiences lower propagation attenuation, which is desired in practical applications. As a consequence, our following discussions are mainly focused on the LRSPhP mode.

The SPhP entangled pairs could be generated in the interaction process of entangled photons and the lossy condensed matter. For a clearer sense, an illustration of the model is presented in Fig. 2. When a pair of entangled photons incident on the condensed matters, they could be transferred into entangled SPhP pairs through surface gratings or the attenuated total reflection (ATR) regime. It is worth mentioning that wavelength of SPhP is usually located in mid-infrared band. Although the usually used entangled photons sources are mostly in near-infrared band, nonlinear optical processes in crystals such as $ZnGeP_2$, CdSe and GaSe could be utilized to provide mid-infrared entangled photon pairs needed to generate entangled LRSPhP pairs[16]. For instance, to prepare a frequency-entangled LRSPhP pair, the corresponding entangled photons could be obtained through the traditional scheme as reported in Ref. 17.

To describe the quantum characters and relative evolution processes of LRSPhP,

a canonical quantization procedure should be taken[18] with the results as,

$$A(r,t)=\sum_k \sqrt{\frac{\hbar}{2\omega_k \varepsilon_0 L^2 |\Gamma_k|}}\left[b_k \Psi_k(x)\exp(i\beta_k z - i\omega_k t) + b_k^\dagger \Psi_k^*(x)\exp(-i\beta_k z + i\omega_k t)\right] \quad (3)$$

,

where the eigenvector $\Psi_k(r)$ could be expressed as

$$\begin{aligned}\Psi_k(x) = \{ &(\frac{i\beta_k}{\alpha_d} x - \hat{z})\exp[\alpha_d(x+h)]\theta(-x-h) + [(\frac{i\beta_k}{\alpha_c}\cdot\frac{\varepsilon_d \alpha_c}{\varepsilon_c \alpha_d} x + \hat{z})\cosh(\alpha_c x) \\ &-(\frac{i\beta_k}{\alpha_c} x + \frac{\varepsilon_d \alpha_c}{\varepsilon_c \alpha_d}\hat{z})\sinh(\alpha_c x)]\theta(-x)\theta(x+h) + (\frac{i\beta_k}{\alpha_d} x + \hat{z})\exp(-\alpha_d x)\theta(x)\}\end{aligned} \quad (4)$$

.

The subscript is marked only by wave number k, as the polariton state of LRSPhP is always TM mode, the usually used subscript σ for polariton could be ignored. $b_k$ represents the annihilation operator of LRSPhP. And $\alpha_q$ is given by $\alpha_q = \left(\beta_k^2 - \frac{\omega_k^2}{c^2}\varepsilon_q\right)^{\frac{1}{2}}, (q = d, c)$. $\Gamma_k$ is the normalization coefficient which could be determined through the normalized longitudinal power flux[19]. Corresponding equation is written as $P_k = \frac{\beta_k}{2\omega_k \varepsilon_0 |\Gamma_k|^2}\int_{-\infty}^{\infty}\frac{1}{\varepsilon_k(x)}|\Psi_k(x)|^2 dx = 1$.

Taken the quantization formulation of the LRSPhP mode, we could obtain the effective interaction Hamiltonian of the coupling process between entangled photons and entangled LRSPhPs. The classical Hamiltonian of the electromagnetic field is written as $H = \frac{1}{2}\int_V \left(\varepsilon_0 \varepsilon(x) E^2 + \mu_0 H^2\right) dv$. To describe the interaction process, the total electric field E in the equation should be expressed as $E = E_{eph} + E_{esp}$, in which $E_{eph}$ and $E_{esp}$ represents the field of entangled optical states and the field of entangled LRSPhPs, respectively. The optical field of entangled photons usually is generated by spontaneous parametric down-conversion (SPDC)[5], so the expression should be obtained from the second order nonlinear polarization $P = \varepsilon_0 \chi^{(2)} : EE$. Taking the actual interaction process into account, the electric field in the expression could be substituted by the $E^{(-)}$ part which is marked by the creation operator of photon $a_k^\dagger$ representing the producing of a photon[1], so the quantized field of the entangled photons could be expressed as follows

$$E_{eph}(r_1, r_2, t) = \sum_k \sum_{(\sigma_1,\sigma_2)} \Re_{12}\Theta(x_1, x_2) a_{k\sigma_1}^\dagger a_{(k_p - k)\sigma_2}^\dagger \exp\{-i[k_z z_1 + (k_{pz} - k_z) z_2 - \omega_p t]\} \quad (5)$$

.

The coefficient $\Re_{12}$ is given by

$$\Re_{12} = -\frac{\hbar\chi^{(2)}}{2\varepsilon_0}\sqrt{\frac{\omega_{k\sigma}(\omega_p - \omega_{k\sigma})}{\Xi_1\Xi_2}} \quad (6)$$

The quantities marked by subscript p are relative parameters of the pump field of the SPDC process, and the perfect phase matching conditions are assumed to be satisfied. $\Xi_1$ and $\Xi_2$ are defined by respective normalization procedures. The eigenvector $\Theta(x_1, x_2)$ could be decomposed into the product of eigenvectors of the two subspaces $\Theta(x_1, x_2) = \vartheta^*(x_1)\vartheta^*(x_2)$ . To investigate the interaction process, the eigenvector $\vartheta(x)$ could be transformed into an equivalent formulation[18.]

$$\vartheta(x) = [\lambda_1 \exp(-\alpha_c x) + \lambda_2 \exp(\alpha_c x)]\theta(-x)\theta(x+h) + \lambda_3 \exp[\alpha_d (x+h)]\theta(-x-h) , \quad (7)$$

where the $\lambda_i$'s (i=1, 2, 3) are determined by the overall boundary conditions of the coupling gratings.

For the entangled LRSPhPs, although the formulation of the field could not be derived through the second order nonlinearity, it is natural to determine that based on the correspondence between frequencies of the incident photons and the LRSPhP modes. The quantized field of the entangled LRSPhP modes thus are be expressed as

$$E_{esp}(r_1, r_2, t) = \sum_{k_1, k_2} \Omega_{12} b_{k_1}^\dagger b_{k_2}^\dagger \exp\{-i[(k_1 z_1 + k_2 z_2) - (\omega_{k_1} + \omega_{k_2})t]\} \quad (8)$$

The coefficient $\Omega_{12}$ is given by

$$\Omega_{12} = -\frac{\hbar}{2\varepsilon_0 L^2}\sqrt{\frac{\omega_{k_1}\omega_{k_2}}{\Gamma_{k_1}\Gamma_{k_2}}}\Psi_{k_1}^*(x)\Psi_{k_2}^*(x) \quad (9)$$

Substituting the formulations of $E_{eph}$ and $E_{esp}$ into the Hamiltonian of the electromagnetic field, the overall interaction Hamiltonian is written as

$$H = \frac{\pi\hbar^2(\chi^{(2)})^2}{2\varepsilon_0}\sum_k \sum_{(\sigma_1,\sigma_2)} I_{k\sigma_1\sigma_2} a_{k\sigma_1}^\dagger a_{(k_p-k)\sigma_2}^\dagger a_{(k_p-k)\sigma_2} a_{k\sigma_1} + \frac{\pi\hbar^2}{2\varepsilon_0 L^4}\sum_{k_1,k_2} I_{k_1k_2} b_{k_1}^\dagger b_{k_2}^\dagger b_{k_2} b_{k_1}$$
$$+\left\{\sum_{k,k_1,k_2}\sum_{(\sigma_1,\sigma_2)} \Im_{kk_1k_2}\mathbb{C}(r_1, r_2) b_{k_1}^\dagger b_{k_2}^\dagger a_{k\sigma_1} a_{(k_p-k)\sigma_2} \exp[-i(\omega_p - \omega_{k_1} - \omega_{k_2})t] + H.C.\right\} , \quad (10)$$

where

$$I_{k\sigma_1\sigma_2} = \frac{\omega_{k\sigma}(\omega_p - \omega_{k\sigma})}{\Xi_1\Xi_2}\int dx_1 dx_2 \varepsilon(x_1)\varepsilon(x_2)\left|\Theta(x_1,x_2)\right|^2$$
$$I_{k_1k_2} = \frac{\omega_{k_1}\omega_{k_2}}{\Gamma_{k_1}\Gamma_{k_2}}\int dx_1 dx_2 \varepsilon(x_1)\varepsilon(x_2)\left|\Psi_{k_1}(x_1)\right|^2\left|\Psi_{k_2}(x_1)\right|^2 \quad (11)$$

In the interaction Hamiltonian, the first and the second terms corresponds to the unperturbed portion of the Hamiltonian. The physical meaning of these two terms is quite clear. Their effective operators are both combinations of the particle number operators for the entangled states, so expectation values of these operators represent the combination detections of entangled photons and entangled LRSPhPs, respectively.

The latter two terms corresponds to the interaction of entangled states. The first one describes the generation of the entangled LRSPhPs, while the latter one represents the conjugate process. So the effective $H_1$ could be written as

$$H_1 = \sum_{k,k_1,k_2}\sum_{(\sigma_1,\sigma_2)} \Im_{kk_1k_2}\mathbb{C}(r_1,r_2) b_{k_1}^\dagger b_{k_2}^\dagger a_{k\sigma_1} a_{(k_p-k)\sigma_2} \exp[-i(\omega_p - \omega_{k_1} - \omega_{k_2})t] \quad (12)$$

Here the coefficient $\Im_{kk_1k_2}$ is given by

$$\Im_{kk_1k_2} = \frac{\hbar^2}{4\varepsilon_0 L^2}\sqrt{\frac{\omega_{k_1}\omega_{k_2}}{\Gamma_{k_1}\Gamma_{k_2}}}\sqrt{\frac{\omega_{k\sigma}(\omega_p - \omega_{k\sigma})}{\Xi_1\Xi_2}} \quad (13)$$

In the derivation, we assume the conservation of energy is satisfied in the SPDC process to prepare the entangled optical fields, so the spatial integral in the interaction Hamiltonian is estimated as

$$\mathbb{C}(r_1,r_2) = \int d^3r_1 \int d^3r_2 h(r_1,r_2)$$
$$=(\int dx_1 dx_2 \varepsilon(x_1)\varepsilon(x_2)\Psi_{k_1}^*(x_1)\Psi_{k_2}^*(x_2)\Theta^*(x_1,x_2))\cdot \sin c(\frac{\Delta\beta_1 z_1}{2})\sin c(\frac{\Delta\beta_2 z_2}{2}) z_1 z_2 \quad (14)$$

To simplify the form of expression, origins of $z_1$ and $z_2$ axes are set in the middle of the gratings, *i.e.*, the coupling sections. $\Delta\beta_1$ and $\Delta\beta_2$ are the relative wave vector mismatches of the coupling processes, $\mathbb{C}(r_1,r_2)$ is plotted as a function of these two quantities as shown in Fig. 3. In the calculations, $z_1$ and $z_2$ are both set at 100 μm. For effective coupling, $\Delta\beta_1$ and $\Delta\beta_2$ should vanish. This condition is stratified by the equivalent reciprocal vector $\Delta\beta_i = \frac{2m\pi}{\Lambda_i}$ provided by the periodical variation of the gratings.

For a more explicit comprehension of the entangled LRSPhP modes, the state vector could be derived from second order perturbation

$$|\Psi\rangle = \left(-\frac{i}{\hbar}\right)^2 \int dt_1 \int dt_2 T[H_{SPDC}(t_1)H_1(t_2)]|0\rangle \tag{15}$$
,

where T is the time-ordering operator. Neglecting the unrelated items, $|\Psi\rangle$ could be written as

$$|\Psi\rangle = -\frac{2\pi}{\hbar^2}\sum_{k_1,k_2} F_k \Im_{kk_1k_2} \mathbb{C}(r_1, r_2)\delta(\omega_p - \omega_{k_1} - \omega_{k_2}) b_{k_1}^{\dagger} b_{k_2}^{\dagger} |0\rangle \tag{16}$$
.

The coefficient $F_k$ is defined by the SPDC process. We would like to take the entangled state for a detailed consideration. As we know, the SPDC-generated entangled photon state is usually entangled in frequency and wave vector[1]. In frequency space, the coupling process from photon to LRSPhP does not influence the corresponding frequency property, so the frequency entanglement derived from entangled photons could completely preserve in the entangled LRSPhPs, which is stressed by the Dirac-δ function in Eq. (19). For the case of wave vector, the situation is slightly different. The wave-vectors along z direction should be considered for the phase-matching condition of the coupling process. Generally speaking, the wave-vector entanglement has implications for the spatial correlations of the entangled pair[1], so the quantum entanglement based on SPhP corresponds to a low dimension spatial correlation system.

For a further consideration, we notice that in the coupling process of entangled photons and entangled LRSPhPs, the former one could not been transformed into the latter one with unity probability. Referring to the classical situation, the coupling coefficient determines the portion of the incident light which could be coupled into LRSPhP mode. For the quantum situation, that would correspond to the transition probability from photons to LRSPhPs. Therefore there are in fact three possible results of the interaction between entangled photons and the grating structures. Besides the transformation from entangled photons to entangled LRSPhPs we have discussed above, it still could generate an entangled pair of photon and LRSPhP, or a surviving entangled photon pair. For an overall treatment of the state transformation including these three processes, the quantum field theory (QFT) method which we use in the above section suffers seriously from its complicacy. As a correspondence, the wave mechanics (WM) method established by Bialynicki-Birula[20] and Sipe[21] may present an intuitive and direct description for the process. In this theory, performances of photons obey the Schrödinger-like Bialynicki-Birula-Sipe equation which is equivalent to the set of Maxwell equations[22].

To investigate variations of two-entity states, the Bialynicki-Birula-Sipe equation should be generalized. The motion equation for the two-photon wave function in vacuum has been derived by Smith and Raymer[22, 23]. We need to expand the equation in dealing with problems in dielectrics. The Bialynicki-Birula-Sipe equation for single photon wave function in dielectrics has been derived in Ref. 24, through a similar treatment; the governing equation for two-state evolution is obtained as

$$i\frac{\partial}{\partial t}\Psi_d^{(2)} = c\alpha_1^{(2)}\nabla_1\times\Psi_d^{(2)} + c\alpha_2^{(2)}\nabla_2\times\Psi_d^{(2)}, \tag{17}$$

where

$$\alpha_1^{(2)} = \Sigma_3\otimes I,\ \ \alpha_2^{(2)} = I\otimes\Sigma_3 \tag{18}$$

with

$$I = \begin{bmatrix}\hat{1} & 0\\ 0 & \hat{1}\end{bmatrix},\ \ \Sigma_3 = \begin{bmatrix}\hat{1} & 0\\ 0 & -\hat{1}\end{bmatrix},\ \ \hat{1} = \begin{bmatrix}1 & 0 & 0\\ 0 & 1 & 0\\ 0 & 0 & 1\end{bmatrix}. \tag{19}$$

The two-state wave function $\Psi_d^{(2)}$ is written as

$$\Psi_d^{(2)}(r_1,r_2,t) = \sum_{(j\sigma),(m\rho)} C_{(j\sigma),(m\rho)}\psi_{j\sigma}^{(1)}(r_1,t)\otimes\psi_{m\rho}^{(1)}(r_2,t). \tag{20}$$

Here (jσ) and (mρ) are subscripts of state 1 and state 2, respectively. The single state wave function in dielectrics is written as

$$\psi^{(1)}(r,t) = \psi_+^{(1)}(r,t) + \psi_-^{(1)}(r,t)$$

$$(\psi_\pm^{(1)}(r,t) = \sqrt{\frac{\varepsilon_0\varepsilon}{2}}E_\pm \pm i\sqrt{\frac{\mu_0}{2}}H_\pm \ \text{ and } \ \sigma\psi_\pm^{(1)} = \pm\psi_\pm^{(1)}) \tag{21}$$

The actual electric field and magnetic field of the system are $E = E_+ + E_-$ and $H = H_+ + H_-$, respectively. It should be emphasized that the photon wave function could not be normalized in the usual sense based on position probability density due to the lack of localizability. As the energy density of the electromagnetic field is localized, it is applied as a substitution in WM of photon, which is equal to the moduli square of Ψ(r, t). For the LRSPP mode, as the energy distribution is inhomogeneous due to the space dependence of permittivity, the corresponding wave function of LRSPhP whose moduli square represents the spacial energy density perhaps possesses potential value to describe the quantum behavior of LRSPhP.

It is noteworthy that these equations are only reasonable for non-absorptive matters. Although the real systems we discussed are lossy, the key process is still the coupling between entangled photons and LRSPhP modes. The propagation attenuation doesn't play a crucial role, thus the imaginary part of the lossy material could be ignored in the derivation. Based on the theories above, the controlling equation of the interaction process is given by

$$\sum_{\nu,\eta} \exp[-i(\Delta\beta_{1,\nu\mu} z_1 + \Delta\beta_{2,\eta\tau} z_2)] F_{\nu\mu}(r_1) F_{\eta\tau}(r_2) = 0 \qquad (22)$$

In the equation, the function $F_{\nu\mu}(r)$ is written as

$$F_{\nu\mu}(r) = (\frac{\partial}{\partial z} - i\Delta\beta_{\nu\mu}) \sum_{(\nu,\mu)} \operatorname{Im}\left\langle \operatorname{Re}\psi_\nu \left| \frac{A}{\sqrt{\varepsilon(x)}} \right| \psi_\mu \right\rangle_S - \frac{i\omega}{c\varepsilon_0} \left\langle \operatorname{Re}\psi_\nu \left| \frac{\Delta\varepsilon}{\varepsilon(x)} \right| \operatorname{Re}\psi_\mu \right\rangle_S \qquad (23)$$

where the operator $A$ is equivalent to $\times\sigma$. And, the function $F_{\eta\tau}(r)$ has a fully corresponding formulation. Through further derivations, the equation could be simplified to

$$\sum_{\nu,\eta} \exp[-i(\Delta\beta_{1,\nu\mu} z_1 + \Delta\beta_{2,\eta\tau} z_2)] \partial'_{z_1} \partial'_{z_2} \xi^{(2)}_{\nu\eta}(z_1, z_2) = 0 \qquad (24)$$

The $\xi^{(2)}_{\nu\eta}(z_1, z_2)$ represents the combination of the longitudinal portion of the wave functions, and the operator $\partial'_{z_j}$ is expressed as

$$\partial'_{z_j} = \lambda_{(\nu\mu),(\eta\tau)} (\frac{\partial}{\partial z_j} - i\Delta\beta_{(\nu\mu),(\eta\tau)}) + i\omega_j \kappa_{(\nu\mu),(\eta\tau)} \qquad (25)$$

$\lambda_{(\nu\mu),(\eta\tau)}$ and $\kappa_{(\nu\mu),(\eta\tau)}$ represents the orthonormalization and coupling coefficients, respectively. Through these functions, the states transformation process is expressed as

$$\begin{bmatrix} \xi_{\nu\eta,f} \\ \xi_{\nu\tau,f} \\ \xi_{\mu\eta,f} \\ \xi_{\mu\tau,f} \end{bmatrix} = \begin{bmatrix} t_{\nu\mu} t_{\eta\tau} & t_{\nu\mu} \kappa_{\eta\tau} & \kappa_{\nu\mu} t_{\eta\tau} & \kappa_{\nu\mu} \kappa_{\eta\tau} \\ -t_{\nu\mu} \kappa^*_{\eta\tau} & t_{\nu\mu} t^*_{\eta\tau} & -\kappa_{\nu\mu} \kappa^*_{\eta\tau} & \kappa_{\nu\mu} t^*_{\eta\tau} \\ -\kappa^*_{\nu\mu} t_{\eta\tau} & -\kappa^*_{\nu\mu} \kappa_{\eta\tau} & t^*_{\nu\mu} t_{\eta\tau} & t^*_{\nu\mu} \kappa_{\eta\tau} \\ \kappa^*_{\nu\mu} \kappa^*_{\eta\tau} & -\kappa^*_{\nu\mu} t^*_{\eta\tau} & -t^*_{\nu\mu} \kappa^*_{\eta\tau} & t^*_{\nu\mu} t^*_{\eta\tau} \end{bmatrix} \begin{bmatrix} \xi_{\nu\eta,i} \\ \xi_{\nu\tau,i} \\ \xi_{\mu\eta,i} \\ \xi_{\mu\tau,i} \end{bmatrix} \qquad (26)$$

In our analysis, the only initial state is the entangled photon state. Through the transformation, four entangled states in three types are generated with corresponding possibilities.

For a clear view, an actual instance of coupling is given in Fig. 4. It is based on SiC systems. To ensure effective transformations, the mismatch of wave vector is analyzed. It could be written as $\Delta\beta = \frac{2\pi}{\lambda}[n_{eff}(\lambda) - \sin\theta] = 2\pi B(\lambda,\theta)$. The function B(λ, θ) is plotted in Fig. 4(a). For the entangled pair, the value of B($\lambda_1$, θ) and B($\lambda_2$, θ) should be the same. Fig. 4(b) shows the coupling efficiency of the situation in which function B is equal to 0.1 ($\theta = \pi/9$), Peak A plus peak B represents the entangled pairs, *i.e.*, ($\omega_1$, $\omega_2$) or ($\omega_2$, $\omega_1$).

To show correlations of the system, the second-order correlation function has been investigated. It could be derived from tracing over the tensor product of the two-photon wave function and its Hermitian conjugate then integrating over all space and dividing by the normalization values. The formulation of the correlation function could be written as

$$g^{(2)}(r_1, r_2; t) = \frac{\iint Tr[\Psi^\dagger(r_1, r_2, t)\Psi(r_1, r_2, t)] d^3r_1 d^3r_2}{\langle I_1 \rangle \cdot \langle I_2 \rangle}, \tag{27}$$

where

$$\langle I_i \rangle = \left( \int \Psi_i^\dagger(r_i, t)\Psi_i(r_i, t) d^3r_i \right) \tag{28}$$

For an unentangled case, the numerator of the correlation function could be decomposed into products of the subspaces. In contrast, the correlation function for entangled states will include coherent superposition of the states which implies the nonlocal correlations. The detailed formulation of correlation function is expressed as

$$g^{(2)}(r_1, r_2; t) \propto F_{\alpha\beta} G(\chi_1, \chi_2), \tag{29}$$

where $F_{\alpha\beta}$ is determined by the specific entangled pair, and ($\alpha$, $\beta$) represents photon or LRSPhP. G($\chi_1$, $\chi_2$) is defined by the specific entanglement scheme. For the energy-time type, $\chi$ represents phase and G($\chi_1$, $\chi_2$) is a sinusoidal function with respect to the phase difference. In the frequency entanglement case, G($\chi_1$, $\chi_2$) is proportional to beating item of the two frequencies.

In summary, the entanglement of SPhPs in condensed matter systems is studied. In comparison with the traditional bulk all-optical quantum circuits, the SPhP based system offers an alternative to realize a much compact integrated quantum optical circuits based on condensed matters. We investigate the quantum performances of entangled LRSPhPs by the interaction Hamiltonian and the perturbation theory. As a comparison, the coupling process is investigated by the wave mechanics approach of quantum theory. Finally, the correlation of system is examined. A whole set of descriptions of SPhP entanglement thus are given.

This work is supported by 973 programs under contract No. 2011CBA00205 and 2012CB921803, the PAPD, Fundamental Research Funds for the Central Universities.

**Figure Captions:**

Fig. 1. (a) Dielectric permittivity curve of SiC; (b) Dispersion curve of SiC, the blue solid lines are those of bulk polaritons, while the red solid line is that of SPhP mode. The oblique dashed line is the light line.

Fig. 2. Illustration of the generation of entangled LRSPhP modes

Fig. 3. Normalized curve of $\mathbb{C}(r_1, r_2)$. Both $z_1$ and $z_2$ are set at 100 μm.

Fig. 4. (a) The values of function B(λ, θ) at different wavelengths and incident angles; (b) The coupling efficiency when B(λ, θ) is equal to 0.1 (θ = π/9). Peak A and B correspond to the transformation from entangled photons to entangled LRSPhP pairs ($\omega_1$, $\omega_2$) + ($\omega_2$, $\omega_1$).

Figure 1

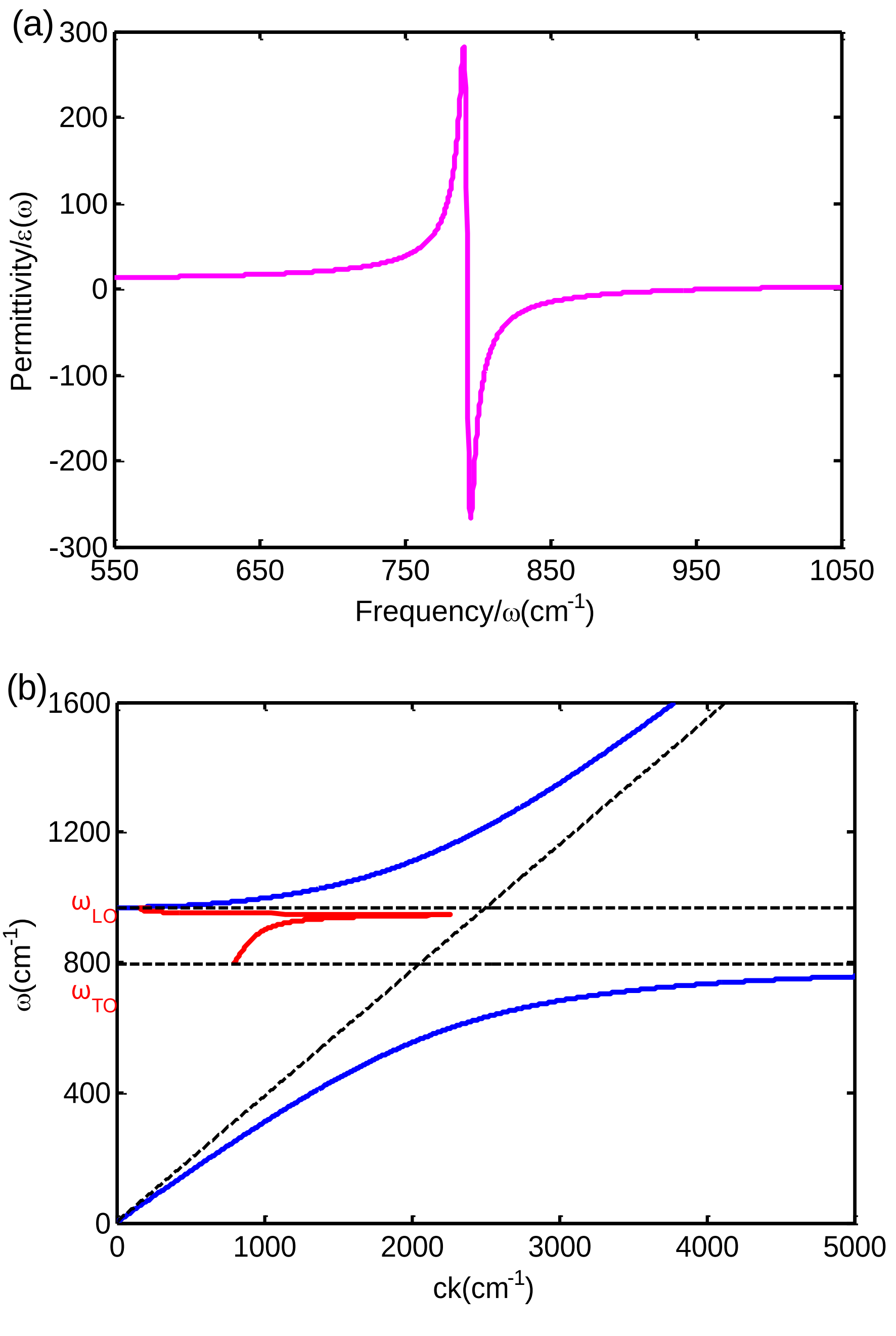

Figure 2

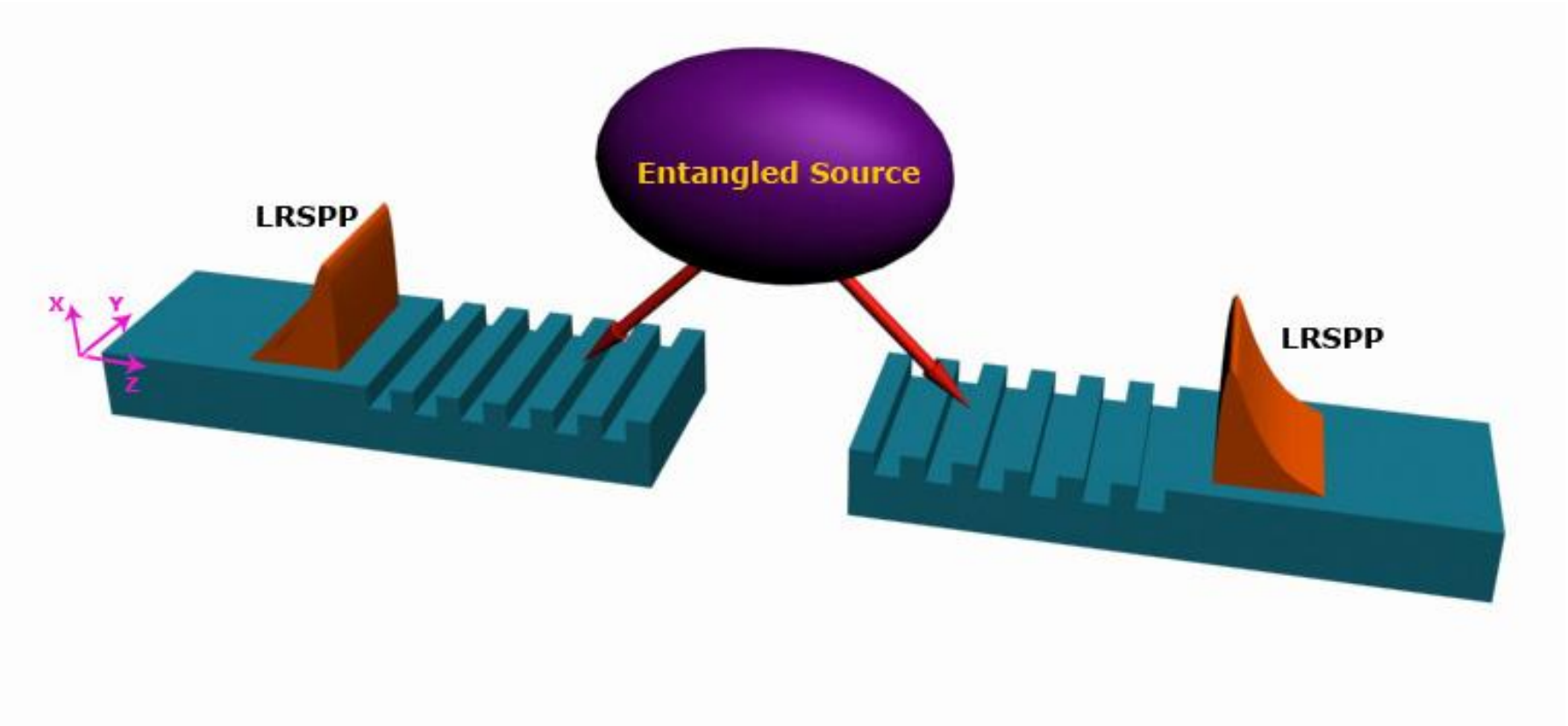

Figure 3

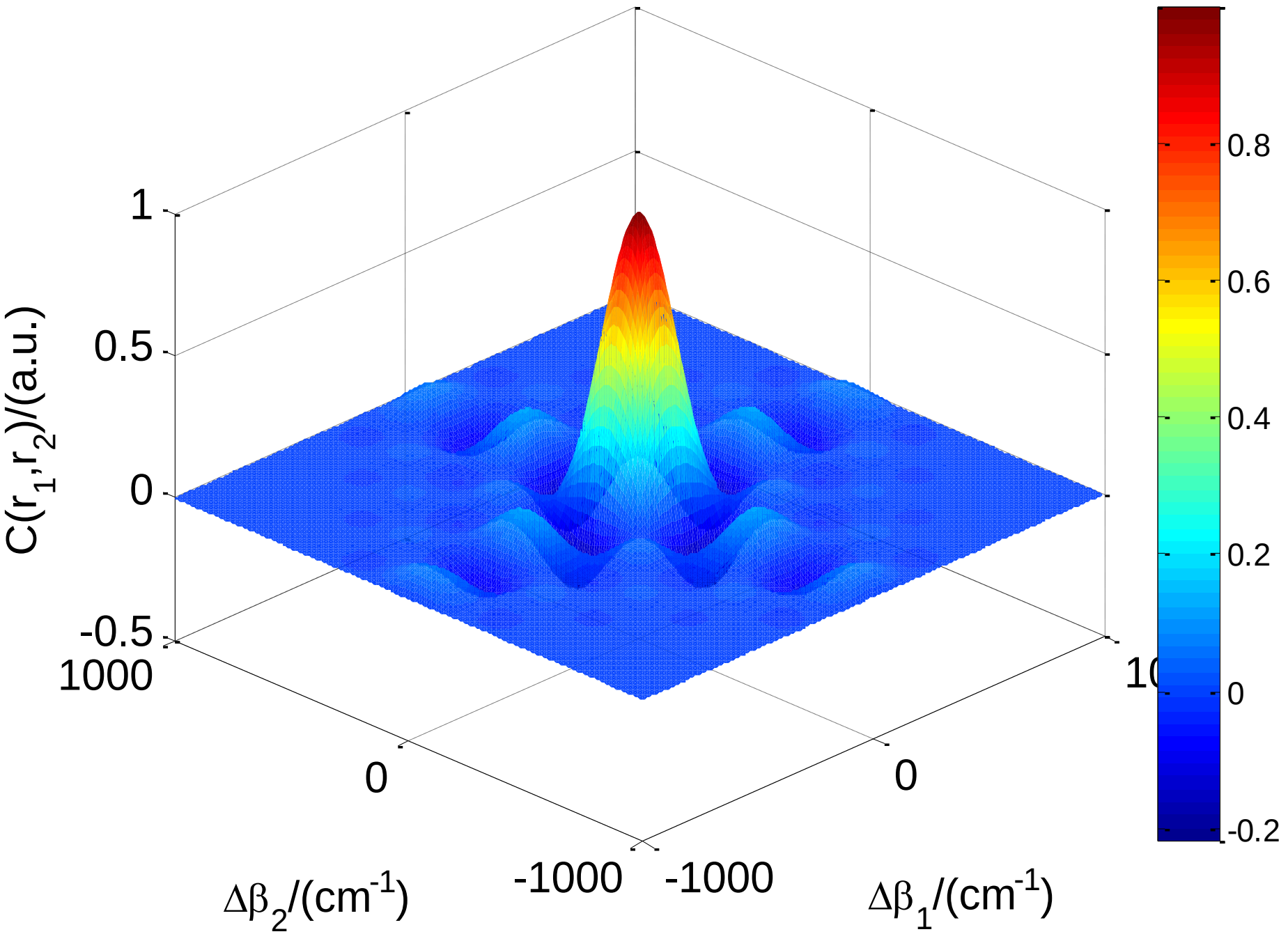

Figure 4

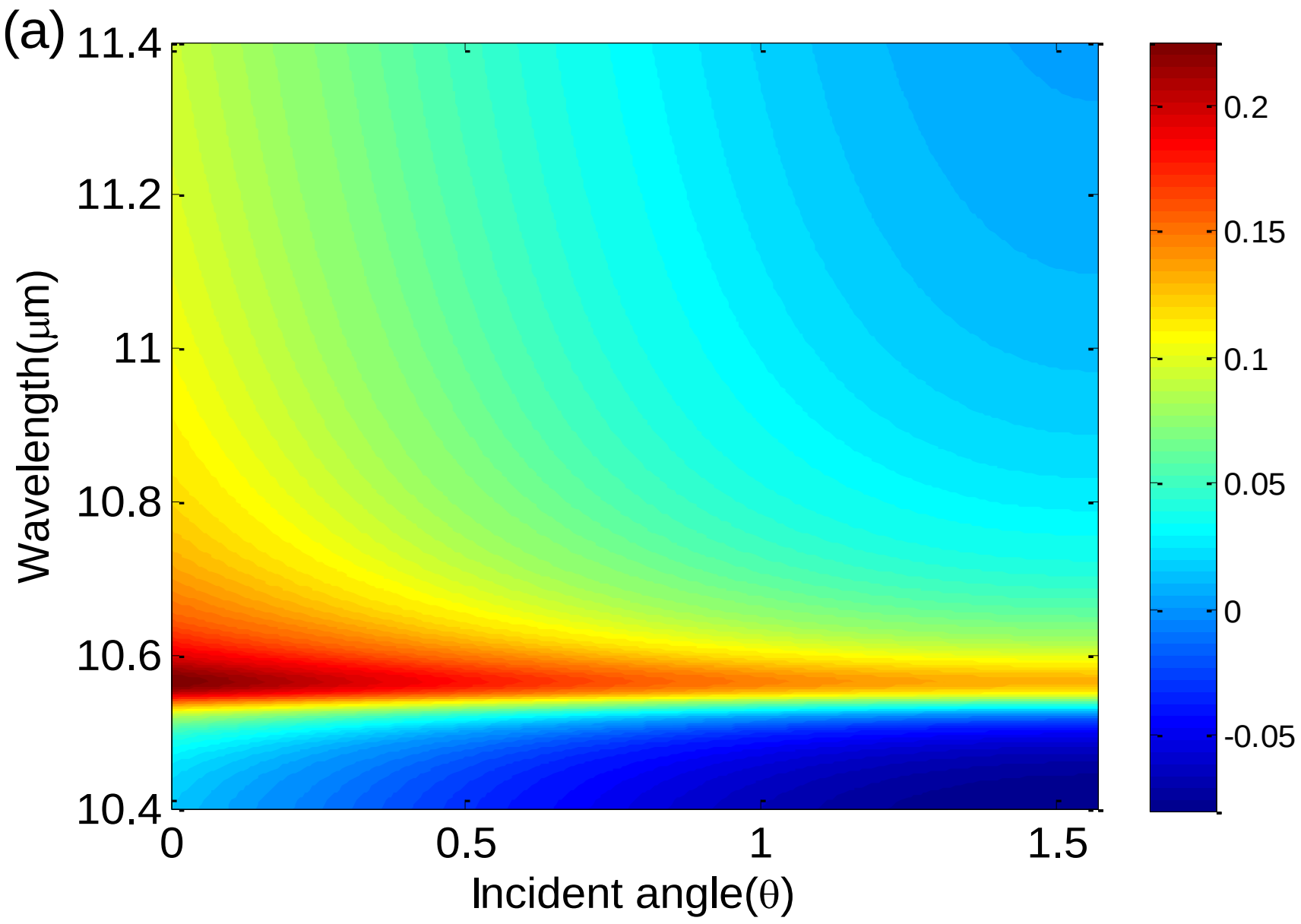

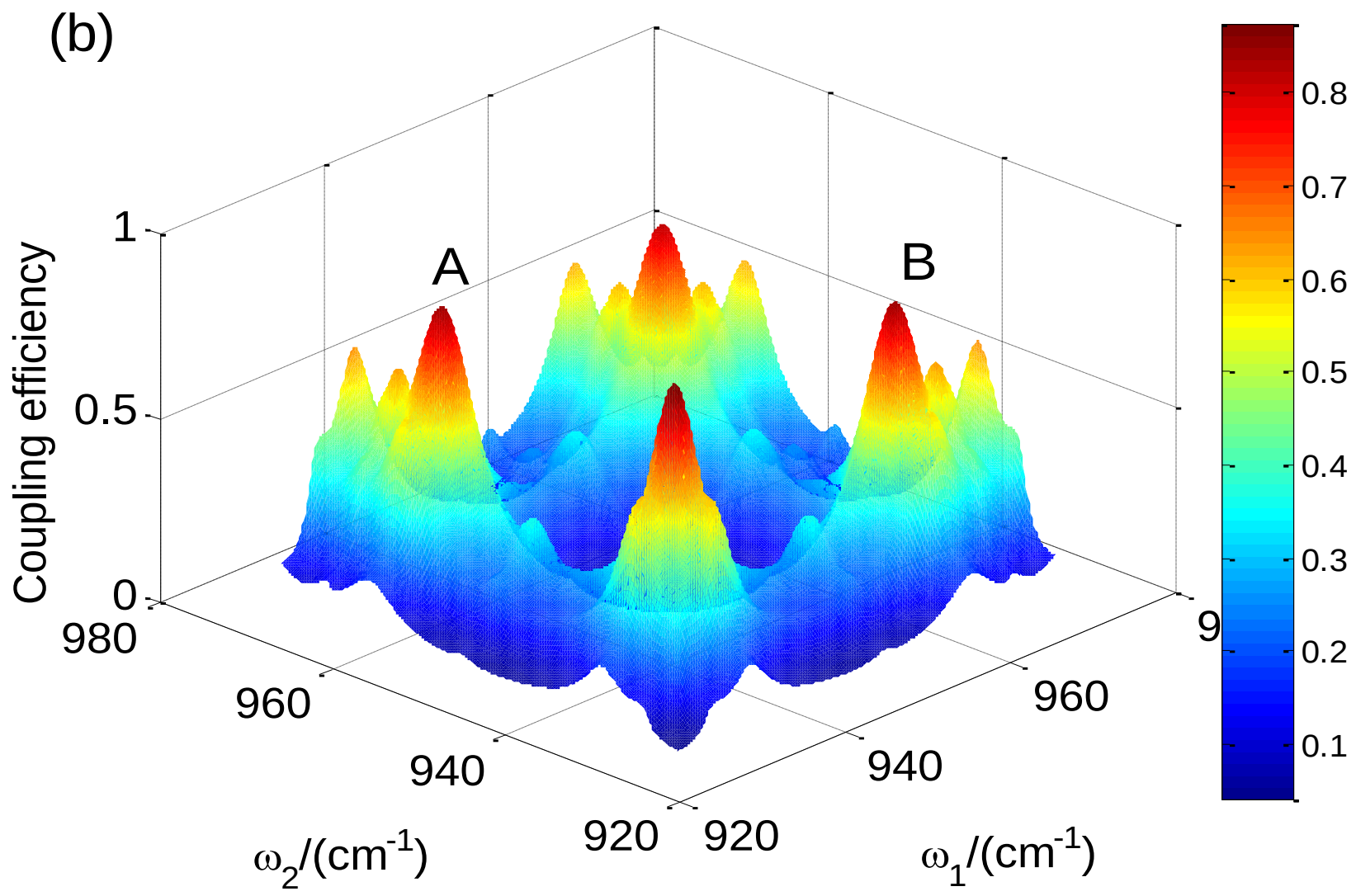